\documentclass[12pt]{iopart}

\usepackage{graphicx}% Include figure files
\usepackage{bm}% bold math
\usepackage{makeidx}
\usepackage{comment}
\usepackage{url}
\usepackage[ansinew]{inputenc}
\usepackage[usenames,dvipsnames]{pstricks}
\usepackage{subfigure}
\usepackage{epsfig}
\usepackage{amssymb}

\begin{document}
\title[Structure preserving noise and dissipation in the Toda lattice]{ Structure preserving noise and dissipation in the Toda lattice}

\author{Alexis Arnaudon$^1$}

\address{$^1$ Department of Mathematics, 
Department of Mathematics, Imperial College, London SW7 2AZ, UK}
\ead{alexis.arnaudon@imperial.ac.uk}
\begin{abstract}
  In this paper, we use Flaschka's change of variables of the open Toda lattice and its interpretation in term of the group structure of the LU factorisation as a coadjoint motion on a certain dual of Lie algebra to implement a structure preserving noise and dissipation. 
  Both preserve the structure of coadjoint orbit, that is the space of symmetric tri-diagonal matrices and arise as a new type of multiplicative noise and nonlinear dissipation of the Toda lattice. 
  We investigate some of the properties of these deformations and in particular the continuum limit as a stochastic Burger equation with a nonlinear viscosity. 
  This work is meant to be exploratory, and open more questions that we can answer with simple mathematical tools and without numerical simulations. 
\end{abstract}
%\submitto{\jpa}

\maketitle

\section{Introduction}

The Toda lattice has its roots in the origin of the theory of integrable systems with the discovery of recurrences in the FPU lattice by \cite{fermi1955studies}, with numerical simulations performed by M. Tsingou, see \cite{dauxois2008fermi} for more on this story.  
The surprise was that for large times, no statistical equilibrium emerged for this lattice despite theoretical predictions of statistical physics. 
It turns out that the FPU chain is an approximation of the Toda lattice, which is an integrable system and has no thermal equilibrium. 
In the FPU chain, if one waits for a long enough time such that the non-integrable effects appear, this expected equilibrium can be observed in precise numerical simulations, see for example \cite{benettin2011time,benettin2013fermi}. 
The Toda lattice was first formulated in \cite{toda1967vibration} (see also \cite{toda1988theory}) and later proved to be integrable by \cite{flaschka1974toda,flaschka1974IItoda} and \cite{henon1974integrals}.
These works triggered numerous studies on such lattices, with various connections to geometry. 

In this work, we want to deform the classical Toda flow, written in a geometrical form, that is as a coadjoint motion on a certain coadjoint orbit, in two different ways. 
First, we will introduce some noise and then a dissipative, or viscous term. 
The form of these terms will not be inspired by any physical processes that can perturb the physical Toda lattice but will be derived from the only condition of preserving the geometry of the Toda flow in Flaschka's variables. 
These deformations seem to have never been studied before, and due to their nonlinear nature, are challenging to study in details. 
This work is thus only trying to open a new door for mathematical investigations on the Toda lattice and can be seen as a collection of open problems. 

\section{The deterministic Toda lattice}\label{toda-lattice}

We begin with a short description of the open Toda lattice to set up our notation and then use Flascka's change of variables \cite{flaschka1974toda} to rewrite the equations of Toda lattice as a coadjoint motion on a particular matrix group. 

\subsection{The open Toda lattice}

The simplest Toda lattice is a one-dimensional chain of $n$ massive particles interacting between nearest neighbours. 
We will denote their positions by $x_i$ and velocities by $y_i$. 
The interaction is exponential, and with constants set to unity, the Toda lattice is described by the Hamiltonian
\begin{equation}
  H_p = \sum_{i=1}^n \frac12 y_i^2 +  \sum_{i=1}^{n-1}  e^{x_{i+1}-x_i}\, .
    \label{Toda-Hamiltonian}
\end{equation}
The open Toda lattice, that is with open endpoints, is then a canonical Hamiltonian system with variables $(x_i,y_i)\in T^*\mathbb R^n$ and Hamiltonian \Eref{Toda-Hamiltonian}. 
The equations of motion are then
\begin{equation}
        \dot x_i  = y_i \, , \qquad  \dot y_i =    e^{x_i-x_{i-1}} - e^{x_{i+1}-x_i}\, ,  \qquad i=1,\ldots n\,,
    \label{Toda-physical}
\end{equation}
where $x_0=-\infty$ and $x_{n+1} = \infty$ for an open, or non-periodic Toda lattice. 
Notice that with an approximation of the exponential interaction by a Taylor expansion up to a cubic order, the Toda lattice reduces to the FPU chain. 

We will not describe more these equations and directly introduce Flaschka's change of variable \cite{flaschka1974toda}, which are given by 
\begin{eqnarray}
    a_i &= \frac12 y_i\,,  \qquad i = 1,\ldots,n\nonumber \\
    b_i &= \frac12  e^{(x_{i+1}-x_i)/2}\,, \ \qquad i=1,\ldots, n-1\, .
    \label{Flaschka-variable}
  \end{eqnarray}
The equations of motion are then 
\begin{eqnarray}
    \dot b_i &= b_i(a_i -a_{i+1})\, ,\qquad i= 1,\ldots, n\nonumber \\
    \dot a_i &= 2(b_{i-1}^2-b_{i}^2)\,, \qquad i=1,\ldots,n-1\,  .
    \label{toda-flaschka}
\end{eqnarray}
The first interest of this simple change of variable is to be able to prove the complete integrability of the Toda lattice using a Lax pair and its associated isospectral problem. 
The second is to unravel interesting geometrical features of the Toda flow.  

We will show the latter in the next section and briefly discuss the Lax pair of the Toda lattice.  
One first constructs the symmetric tri-diagonal matrix $L$ with the $a_i$ variables on the diagonal the $b_i$ variables on the first upper and lower diagonal. 
We then define the corresponding $M$ matrix as 
\begin{equation}
    M= L^+-L^-\, , 
\end{equation}
where $L^+$ is the upper triangular part of the $L$ and $L^-$ its lower triangular part. 
Integrability is then a consequence of being able to rewrite the Toda \Eref{toda-flaschka}, and thus \Eref{Toda-physical} as the Lax equation \cite{lax1968integrals}
\begin{equation}
    \dot L = [L, M]\, , 
  \label{Lax-Toda}
\end{equation}
where the commutator is the standard matrix commutator. 
The proof of integrability then relies on using the isospectral problem associated with the operator $L$, with time-dependence of the spectral data given by the operator $M$. 
We refer to \cite{ablowitz1981solitons} for more details on this theory and to \Sref{isospectral-toda} where we will introduce the stochastic deformation of this spectral problem. 
We now introduce more geometry to understand the Lax \Eref{Lax-Toda} as a coadjoint motion on the dual of a certain Lie algebra. 

\subsection{Geometry of the Toda lattice}
For the description of the eometry of the Toda lattice in the context of the Adler-Konstant-Symes theory \cite{adler1978trace,symes1980systems,kostant1979solution}.  We will mostly follow the exposition of \cite{deift1989matrix,deift1986toda}, but we refer the reader to the book \cite{perelomov1990integrable} for more details. 
We also want to refer to the recent work of \cite{bloch2017geometric} based on this geometric interpretation where the authors showed that this change of variables is the inverse of a particular momentum map. 

The geometry of the Toda lattice arises from the LQ decomposition of a matrix $g\in GL(n)$, that is a matrix with non-vanishing determinant, taken only positive here. 
The LQ decomposition is $g= g_Q g_L$, where $g_Q\in SO(n)$ and $g_L\in L(n)$. 
We denote by $L(n)$ the group of lower triangular $n$ by $n$ matrices and by $SO(n)$ the special orthogonal group of dimension $n$.
Using this decomposition, we can define a group multiplication in the space of matrices with strictly positive determinant as $h*g:= g_Qh g_L$. 
This group multiplication makes this space a Lie group, that will be denoted by $G_{QL}$. 
In order to construct the Lie bracket of the associated Lie algebra $\mathfrak g_{QL}$ of this group, we will need the isomorphism 
\begin{eqnarray*}
m: G_{QL}&\to& SO(n)\times L(n)\\ 
g &\mapsto& (g_Q^{-1}, g_L)\,. 
\end{eqnarray*}
With this decomposition, we can define two particular projections.  
One on $\mathfrak l$, the Lie algebra of $L(n)$ and the other on $\mathfrak q$, the Lie algebra of $SO(n)$.  
They are given as
\begin{equation}
    \pi_\mathfrak{q} \xi = \xi_+-\xi_+^T \quad\mathrm{and}\quad  \pi_\mathfrak{l}\xi = \xi_- + \xi_0 + \xi_+^T\, ,
\end{equation}
for any element $\xi\in \mathfrak{g}_{QL}$ and where the subscripts $+$ stands for the strict upper triangular part of $L$, the subscript $-$ the strict lower triangular part of $L$ and the subscript $0$ the diagonal part of $L$. 
Finally, the Lie bracket on $\mathfrak{g}_{QL}$ is given by 
\begin{equation}
  [\xi, \eta]_{\mathfrak g_{QL}} = [\pi_\mathfrak{l} \xi, \pi_\mathfrak{l} \eta]-  [\pi_\mathfrak{q} \xi, \pi_\mathfrak{q} \eta]\qquad \mathrm{for}\  \xi,\eta \in \mathfrak{g}_{QL}\, . 
    \label{QL-bracket}
\end{equation}
With these projections, we can also define the adjoint action of the group $G_{QL}$ on its Lie algebra or its dual, as 
\begin{eqnarray}
    \mathrm{Ad}_g(\xi) &= g_Q^{-1} (\pi_\mathfrak{q} \xi )g_Q + g_L^{-1} (\pi_\mathfrak{l} \xi )g_L \\
    \mathrm{Ad}^*_{g^{-1}}(\mu) &= \pi_{\mathfrak {l}^\perp} g_Q^{-1} \mu g_Q +\pi_{\mathfrak{q}^\perp} g_L^{-1}  \mu g_L \, , 
\end{eqnarray}
where $\xi\in \mathfrak{g}_{QL}$ and $\mu\in \mathfrak{g}^*_{QL}$. 
For the coadjoint action, we used the projection onto the transposed of the Lie algebras $\mathfrak l$ and $\mathfrak q$, which are defined  using the trace pairing as
\begin{equation}
    \langle \xi, \eta\rangle = \mathrm{Tr}(\xi^T\eta)\, . 
\end{equation}
These two other projections are thus 
\begin{equation}
  \pi_{\mathfrak{l}^\perp} \xi = \xi_+-\xi_-^T \quad\mathrm{and}\quad  \pi_{\mathfrak{q}^\perp} \xi = \xi_- + \xi_0\, +\, \xi_-^T\, .
\end{equation}

A convenient way to write the Lie bracket, and its dual, is to use the classical $R$-matrix \cite{semenov1983classical} associated to this decomposition of the Lie algebra $\mathfrak{g}_{QL}$. 
It is defined as 
\begin{equation}
    R= \frac12 ( \pi_\mathfrak{q}- \pi_\mathfrak{l})= \pi_\mathfrak{q}- \frac12\,,
    \label{R-toda}
\end{equation}
and allows us to rewrite the Lie bracket of \Eref{QL-bracket} as the $R$-bracket  
\begin{equation}
    \mathrm{ad}^R_\xi \eta := [\xi, \eta]_R := [R\xi, \eta] + [\xi, R\eta]= [\xi, \eta]_{\mathfrak g_{QL}}\, ,    
    \label{ad-R}
\end{equation}
where we defined the notation $\mathrm{ad}^R$ for the adjoint action of the Lie algebra $\mathfrak{g}_{QL}$ on itself with respect to the $R$- matrix. 
We can also define the adjoint of the $R$-matrix from the trace pairing
\begin{equation}
    \langle \xi, R\eta\rangle =: \langle R^*\xi, \eta\rangle\, , 
\end{equation}
or in term the trace of matrices 
\begin{equation}
    \mathrm{Tr}( \xi^T R\eta) = \mathrm{Tr}( \eta^T R^T\xi )\, ,
\end{equation}
so that we explicitly have
\begin{equation}
    R^*= R^T= \frac12 ( \pi_{\mathfrak{l}^\perp}- \pi_{\mathfrak{q}^\perp})= \pi_{\mathfrak{q}^\perp}- \frac12\, ,
\end{equation}
because $\mathfrak l^\perp$ is adjoint of $\mathfrak q$ and $\mathfrak q^\perp $ is adjoint of $\mathfrak l$ under the trace pairing. 

Using the $R$-bracket and its dual, we can write the coadjoint action as the dual of the adjoint action in \Eref{ad-R} in the trace pairing as 
\begin{equation}
    \mathrm{ad}^{R,*}_\xi\mu = [(R\xi)^T,\mu] + R^T[\xi^T,\mu]\, ,
    \label{coadjoint-R}
\end{equation} 
for $\xi\in \mathfrak g_{QL}$ and $\mu\in \mathfrak g_{QL}^*$. 

From this coadjoint action, there is $R$-Lie-Poisson bracket, a Poisson bracket linear in the momentum that we now write $L\in \mathfrak{g}_{QL}$, of the form  
\begin{equation}
        \{F,G\}(L) = \mathrm{Tr}\left (L^T\left [\frac{\partial F}{\partial L},\frac{\partial G}{\partial L}\right ]_R\right )= \mathrm{Tr}\left (\mathrm{ad}^{R,*}_{\frac{\partial F}{\partial L}}L \frac{\partial G}{\partial L}^T\right )\, .
    \label{R-LP-bracket}
\end{equation}
From this Lie-Poisson bracket, one can directly see that all the traces of powers of the momenta $L$, that is the functions  
\begin{equation}
    H_i = \frac1i \mathrm{Tr}(L^i)\, , 
\end{equation}
are commuting integrals of motion.  
Notice that for a Toda lattice of $n$ particles, only $n$ integrals of motion are independent, and therefore useful for the proof of integrability. 
They correspond to the integral of motion derived by \cite{henon1974integrals}, after applying the inverse of Flaschka's change of variables.
For $i=1$, the integral is the total momentum, and for $i=2$, it is the energy or the Hamiltonian for the Toda flow. 
Indeed, the Lie-Poisson equation or coadjoint motion for any Hamiltonian $H_i$ is given by
\begin{equation}
  \dot L =  \mathrm{ad}^{R,*}_\frac{\partial H_i}{\partial L} L = \left [\left (R\frac{\partial H_i}{\partial L}\right )^T,L\right ] + R^T\left [\frac{\partial H_i}{\partial L}^T,L\right ]\, .
  \label{Toda-LP}
\end{equation}
For the Hamiltonian $H_2 = \frac12 \mathrm{Tr}(L^2)$, we have $\frac{\partial H_1}{\partial L} = L$, which is a symmetric matrix, so the second term of the Lie-Poisson bracket vanishes and the Toda flow becomes
\begin{equation}
    \dot L = [(RL)^T,L ]= [(\pi_\mathfrak{q} L)^T,L]=  -[\pi_\mathfrak{q} L,L]\, . 
\end{equation}
This last equality corresponds to the Lax \Eref{Lax-Toda}, where $M= \pi_\mathfrak{q} L$ only because the second term of \Eref{Toda-LP} vanishes. 

\section{The stochastic Toda lattice}\label{stochastic-toda}

In this section, we will introduce a stochastic deformation of the Toda lattice equation which preserves the geometrical structure described in the previous section, that is the fact that the Toda flow is a certain coadjoint motion. 
We will first review how to do such deformation in general, and then apply it to the Toda flow. 
We refer to \cite{holm2015variational,arnaudon2016noise,cruzeiro2017momentum} for more details and other application of this structure preserving stochastic deformation of dynamical systems.

\subsection{Stochastic coadjoint motion}

In the previous section, we rewrote the Toda lattice equations as a coadjoint motion on the dual of the Lie algebra $\mathfrak g_{QL}$. 
Such dynamical systems can be obtained via a reduction by the symmetry of a motion on the Lie group $G_{QL}$, when the Lagrangian, or Hamiltonian is invariant with respect to the group action, or multiplication in this case. 
In order to arrive at the Toda equation, the group multiplication induced by the QL decomposition must be used. 
One can use other matrix decomposition to obtain other similar coadjoint motions on a different Lie group, see for example \cite{deift1989matrix,deift1986toda}.

Here, we want to write a stochastic differential equation which is compatible with the reduction by symmetry in the sense that the stochastic process is a coadjoint motion and thus preserves the coadjoint orbit, the orbit in the group in the dual of its Lie algebra under the coadjoint motion. 
There are several ways of introducing such a structure preserving noise in any coadjoint motion, but the most fundamental one is to do it in the reconstruction relation, which related the solution on the dual of the Lie algebra to the solution on the Lie group, that is the relation $\dot g = g\xi$, where the multiplication on the right-hand side stands for the tangent lift of the group action to the tangent space.
From this equation, the theory of reduction by symmetry states that the dynamical equation for the momentum variable conjugate to $\xi$ is the Euler-Poincar\`e, or Lie-Poisson equation, see \cite{marsden1974reduction,marsden1999book} for more details on this theory. 

In this work, the noise will be a set of $k$ filtered Wiener processes $W_t^l$ with filtrations $\mathcal F_t^l$. 
See \cite{oksendal2003stochastic} for the definitions of Wiener processes, or Brownian motions with filtrations. 
Associated with these processes, we fix $k$ Lie algebra elements $\sigma_l$ and perturb the reconstruction relation as follow
\begin{equation}
  dg = g\xi dt + \sum_{l=1}^k g\sigma_l\circ  dW_t^l\, , 
\end{equation}
where $\circ$ stands for Stratonovich integration, which makes the normal rule of calculus also valid in this stochastic context.  
From this reconstruction relation, one should use Hamilton-Pontryagin principle to derive the stochastic Euler-Poincar\'e equation as a stochastic coadjoint motion associated with this stochastic reconstruction relation, see \cite{arnaudon2016noise,cruzeiro2017momentum} or \cite{arnaudon2017stochastic}. 
We will not show this derivation here but only state the result, which is that for any Lie group and reduced Hamiltonian $h:\mathfrak g^*\to \mathbb R$, the stochastic Lie-Poisson equation for the momentum $\mu\in \mathfrak g^*$ is 
\begin{equation}
  d \mu = \mathrm{ad}^*_{\frac{\partial h}{\partial \mu}} \mu dt + \sum_{l=1}^k \mathrm{ad}^*_{\frac{\partial \Phi_l}{\partial \mu}} \mu \circ dW_t^l\, ,
  \label{sto-EP}
\end{equation}
where $\Phi_l(\mu) = \langle \sigma_l, \mu\rangle$ are the stochastic potentials. 
From the form of this equation, it is now clear that the coadjoint orbits are preserved by this stochastic flow, but not the Hamiltonian $h$, unless $h$ and $\Phi_l$ are commuting integrals of the deterministic system.

\subsection{Stochastic Toda lattice}

We will not reproduce the previous theory for the Toda lattice but only uses its main result, the stochastic Euler-Poincar\'e \Eref{sto-EP}.
The drift is the Toda flow defined previously for the $L$ matrix and Hamiltonian $H_2$, and the stochastic potentials are taken here to be $\Phi_l = \mathrm{Tr}\left ( L^T \sigma_l\right ) $, where the set of $\sigma_l$ span the symmetric tri-diagonal matrices with zeros on the diagonal, so $k=n$, the number of particles.
Other choices could be made, with for example less $\sigma_l$ fields and noise only on specific sets of particles, but this choice is the simplest and can be made uniform by setting $\|\sigma_l\|=\sigma$ for all $l$. 
Also, any non-vanishing terms on the diagonal will not appear in the equation of motion, because of the structure of the coadjoint motion.

From the stochastic Euler-Poincar\'e \Eref{sto-EP}, we have the stochastic Toda lattice equation 
\begin{equation}
  dL = -\mathrm{ad}^{R,*}_L Ldt + \sum_{l=1}^k\mathrm{ad}^{R,*}_{\sigma_l} L \circ dW_t^l\, ,
\end{equation}
and using the definition of the coadjoint actions in term of the $R$-matrix the equation becomes
\begin{equation}
  dL = -\left [\pi_\mathfrak{q} L ,L\right ] dt + \sum_{l=1}^k  ([(R\sigma_l)^T,L] + R^T[\sigma_l,L])\circ dW_t^l\, .
    \label{Toda-sto}
\end{equation}
By construction, this stochastic flow preserves the coadjoint orbit, namely the tri-diagonal symmetric matrices where $L$ is defined. 
This means than the elements of $L$ that are always $0$ in the deterministic Toda flow remain $0$ in this stochastic flow, independently on the realisation of the noise or the choice of $\sigma_l$.
This fact can be directly verified by computing the stochastic equation in term of the variables $a_i$ and $b_i$, to get
\begin{eqnarray}
    \dot b_i &= b_i(a_i -a_{i+1}) \label{TodaS-Fb} \\
    da_i &= 2(b_{i-1}^2-b_{i}^2)dt +2 \sigma_{i-1} b_{i-1}dW_t^{i-1}- 2\sigma_i b_idW_t^i\, , \label{TodaS-Fa}
\end{eqnarray}
which does not contain any other non-vanishing terms on the right hand side that the diagonal in \Eref{TodaS-Fa} and the first diagonal in \Eref{TodaS-Fb}.
Notice that only the $a$ equation has noise terms, thus the It\^o correction term vanishes and the It\^o or Stratonovitch interpretation of the stochastic integral have the same form, and we will write everything with It\^o integrals, that is without the symbol $\circ$. 
We also used the indices $i$ for the noise as well as the particles because we chose $\sigma_l=\sigma_i$ such that it non-vanishing only in the $i$ entry. 

Using Flaschka's change of variable \Eref{Flaschka-variable}, the stochastic Toda lattice equations can be written in term of the physical variables $(x_i,y_i)$ 
\begin{eqnarray}
        \dot x_i & = y_i \\
        dy_i &=   \left ( e^{x_i-x_{i-1}} - e^{x_{i+1}-x_i}\right )dt +2\sigma_{i-1} e^{(x_i-x_{i-1})/2}dW_t^{i-1}\nonumber \\
    &-2\sigma_i e^{(x_{i+1}-x_i)/2} dW_t^i\, ,
    \label{TodaS}
\end{eqnarray}
which is in fact a type of Langevin equation with multiplicative noise as the momentum variables $y_i$ can be absorbed to get
\begin{eqnarray}
        \ddot x_i &=  e^{x_i-x_{i-1}} - e^{x_{i+1}-x_i} \\
        &+ 2\sigma_{i-1} e^{(x_i-x_{i-1})/2}\dot W_t^{i-1}- 2\sigma_i e^{(x_{i+1}-x_i)/2} \dot W_t^i\, ,
    \label{TodaS-langevin}
\end{eqnarray}
where we formally denote the Brownian motion by $\dot W_t^i$. 
In these variables, the Langevin \Eref{TodaS-langevin} can be formulated in the stochastic variational principle \cite{bismut1982mecanique}
\begin{equation}
    S =  \int\sum_i \left( \dot x_i ^2 -  e^{x_{i+1}- x_i}\right )  dt  + \int \sum_i 4 \sigma_ie^{(x_{i+1}- x_i)/2}\circ dW_t^i\, ,  
\end{equation}
for free variations of $x_i$ and $\dot x_i$. 

The stochastic \Eref{TodaS} are also Hamiltonian with Hamiltonian 
\begin{equation}
  H = \frac12 \sum_i y_i^2 - e^{x_{i+1}- x_i}\, , 
  \label{Toda-ham}
\end{equation}
the stochastic potentials 
\begin{equation}
    \Phi_i =  4 \sigma_i e^{(x_{i+1}-x_i)/2}\, .
\end{equation}
Because these potential do not depend on the velocity, the stochastic terms are only on the momentum equation, that is, the stochastic Hamilton's equations are  
\begin{equation}
    d x_i = \frac{\partial H_1}{\partial y_i} dt\, ,\qquad d y_i = -\frac{\partial H_1}{\partial x_i} dt-\sum_i\frac{\partial \Phi_i }{\partial x_i}\circ dW_t^i\, .
\end{equation}
Interestingly, it is not possible to add any noise in the equation for the position with this derivation, whereas a similar derivation, but for a completely different system of interacting particles can introduce a noise in the position equation. 
This is the case for the stochastic deformation of landmarks dynamic, as singular solutions of a partial differential equation, see for example \cite{arnaudon2016stochastic}. 
The form of the noise is thus tied to the structure of the coadjoint motion of the equations when the stochastic potential is linear in the momentum (which is not $y_i$ in this geometrical setting, but the tri-diagonal matrix $L$). 

\subsection{Conservation laws}

From the equations of motion \Eref{TodaS}, it is simple to see that $H_1= \sum_i y_i$ remains an exact conserved quantity even in presence of noise but that the other deterministically conserved quantities $H_i$ for $i>1$ are not conserved by the stochastic flow. 
This is to be expected as the noise preserves the coadjoint orbit, and the quantities $H_i$ are not Casimirs on this orbit, but only dynamically conserved quantities. 
To get more insights into this, we compute their time evolution to obtain
\begin{eqnarray*}
    \frac{1}{k+1}dL^{i+1}&=& \langle L^k, \mathrm{ad}^*_{\sigma_i}L\rangle \circ dW_t^i\\
    &=& \frac12 \langle\mathrm{ad}_{\sigma_i} L^k, \mathrm{ad}^*_{\sigma_i} L\rangle dt  + \frac k2 \langle\mathrm{ad}_{\sigma_i} (L^{k-1}\mathrm{ad}^*_{\sigma_i}L) , L\rangle  dt + \langle\mathrm{ad}_{\sigma_i} L^k, L\rangle  dW_t^i\\
    &=& \frac k2 \langle L^{k-1}\mathrm{ad}^*_{\sigma_i}L , \mathrm{ad}^*_{\sigma_i} L\rangle  dt + \langle\mathrm{ad}_{\sigma_i} L^k, L\rangle  dW_t^i\, .
\end{eqnarray*}
For the Toda Hamiltonian, that is for $k=1$, its expected derivative satisfies  
\begin{equation}
     \mathcal D_L H_2 =\frac12 \|\mathrm{ad}^*_{\sigma_i}L \|^2\, , 
\end{equation}
where $\mathcal D_L$ can also be seen as the infinitesimal generator of the stochastic process $L$. 
Its exact definition is, for any function $f$ of the stochastic process, 
\begin{equation} 
    \mathcal D_L f(L(t)):= \lim _{\Delta t\to 0 } \mathbb E \left (\left . \frac{f(L({t+\Delta t})- f(L(t))}{\Delta t}\right | \mathcal F_t\right )\, ,
    \label{D-def}
\end{equation}
where $\mathcal F_t$ is the filtration of the stochastic process, see \cite{oksendal2003stochastic,zambrini2015research,arnaudon2017stability} for more details on this derivative. 

If the noise is uniform, that is $\|\sigma_i\|= \sigma$ for all $i$, we can go further to find that  
\begin{equation}
    \mathcal D_LH_2(L(t)) \leq 4\sigma^2 \mathbb E(V(L(t))|\mathcal F_t)\, , 
    \label{DH2}
\end{equation}
where $V$ denotes the deterministic potential energy, that is the exponential part of the Hamiltonian in \Eref{Toda-ham}. 
This formula can be computed explicitly, from the It\^o correction term
\begin{eqnarray*}
    \sum_i y_i d y_i &= \sum_i 2\sigma y_i e^{(x_i-x_{i-1})/2}\circ dW_t^{i-1}-\sum_i 2\sigma y_i  e^{(x_{i+1}-x_i)/2}\circ  dW_t^i \\
        &= \sum_i 2\sigma y_i e^{(x_i-x_{i-1})/2} dW_t^{i-1}-\sum_i 2\sigma y_i  e^{(x_{i+1}-x_i)/2}  dW_t^i \\
        &+ 2 \sigma^2 \sum _i\left ( e^{x_i-x_{i-1}} +e^{x_{i+1}-x_i}\right ) dt- 4\sigma^2\sum_i  e^{(x_{i+1}- x_{i-1})/2}dt \, , 
\end{eqnarray*}
where the result is obtained after discarding the negative and stochastic terms in the last equality.  
We can then integrate \Eref{DH2} in time to obtain a time-dependent upper bound for the mean energy
\begin{equation}
       \mathbb E(H_1(L(t)))\leq H_1(L(0)))\, e^{4\sigma^2 t}\, . 
\end{equation}
This computation can be done using the probability density of the process, see for example \cite{arnaudon2017stability}. 

This bound shows that the effect of the noise in this system is to introduce some energy into the system which can become unbounded as time tends to infinity. 
This property suggests that we would need a dissipative mechanism that would balance this energy input. 
This will be done in the next section \ref{dissipative-toda}, but before we want to address the question of complete integrability of this stochastic Toda lattice. 

\section{The isospectral stochastic Toda lattice}\label{isospectral-toda}

The Toda lattice is the first example of an integrable chain of interacting particles which admits solitons solutions and can be integrated via the theory of inverse scattering transform, after Flaschka's change of variables. 
In the present context of structure preserving stochastic deformation of an integrable system, the question of integrability of the resulting equation is interesting to ask.  
For lower dimensional integrable systems such as the free rigid body or the Lagrange top, we found in \cite{arnaudon2016noise} that this type of stochastic perturbations only preserves integrability if the noise fields $\sigma_l$ have a particular form, compatible with the original deterministic system. 
In fact, one must have that the stochastic potentials $\Phi_l(\mu) = \langle \sigma_l, \mu\rangle$ must be commuting integrals of the deterministic flow. 
For the rigid body, that is the case if the moment of inertia has a cylindrical symmetry, and if $\sigma_l$ is aligned with this axis of symmetry. 
A similar condition holds for the Lagrange top, which has a cylindrical symmetry by construction. 

Interestingly, such a simple integrable reduction does not exist for the stochastic Toda lattice, except if one uses a single Wiener process, with $\sigma_i$ the identity on the first diagonals. 
Such a noise would correspond to a stochastic total momentum, as in the simplest case of the stochastic integrable AKNS hierarchy studied in \cite{arnaudon2015integrable}.
Nevertheless, it is possible to find an isospectral stochastic deformation of the Toda lattice, but losing Flaschka's change of variable, thus a physical interpretation. 
This is achieved by simply removing the term of the noise proportional to $R^*$ in the stochastic Toda lattice \Eref{Toda-sto}, to obtain 
\begin{equation}
    dL = -\left [\pi_\mathfrak{q} L ,L\right ] dt + \sum_i  [(R\sigma_i)^T,L] \circ dW_t^i\, .
    \label{TodaS-iso}
\end{equation}
This equation is isospectral in the following sense. 
If one introduced the spectral problem for the time-dependent vector $\psi$, 
\begin{eqnarray}
    L\psi &= \lambda \psi\\    
    d\psi &= -\pi_\mathfrak{q}L\psi dt + \sum_i (R\sigma_i)^T\psi \circ dW_t^i\, ,
    \label{Sto-iso}
\end{eqnarray}
where the first line is the spectral problem and the second line corresponds to the time evolution of the eigenfunction $\psi$, and verifies that the spectrum is time-independent only if $L$ satisfies the stochastic Toda lattice \Eref{TodaS-iso}. 

This is indeed the case, as one can see by first computing
\begin{eqnarray*}
    d(L\psi) &= dL \psi + L d\psi\\
    &= dL \psi - L \pi_\mathfrak{q}L\psi dt + L \sum_i (R\sigma_i)^T\psi \circ dW_t^i\, ,
\end{eqnarray*}
then
\begin{eqnarray*}
    d(\lambda \psi)&= d\lambda \psi + \lambda d\psi \\
    &= -\lambda  \pi_\mathfrak{q}L\psi dt + \lambda \sum_i (R\sigma_i)^T\psi \circ dW_t^i\\ 
    &=   -\pi_\mathfrak{q}L L \psi dt +  \sum_i (R\sigma_i)^T L\psi \circ dW_t^i\, ,
\end{eqnarray*}
where we used $L\psi= \lambda\psi$ for the last equality. 
The isospectrality comes from the fact that these two computations are equal, so we collect terms and rearrange them to obtain the stochastic Toda equation \Eref{TodaS-iso}.

One can directly check that the original conserved quantities $H_i$ remain conserved for all $i$ by this stochastic flow, but the coadjoint orbit is not preserved, as the matrix $L$ is a full matrix for any tri-diagonal initial conditions. 
The connection with the physical Toda lattice via Flaschka's change of variables is thus lost. 

Although this equation has a stochastic isospectral problem which should be solvable, the question of integrability from Liouville's theorem\cite{arnold89mechanics} remain as the fact that $L$ is a full matrix introduces more degrees of freedom, thus more commuting integral of motions are needed for Liouville's integrability. 
We will not try to solve this problem here but only mention that in \cite{deift1989matrix}, a generalised Toda flow on the coadjoint orbit of full symmetric matrices was proven to be completely integrable, and the formulae for the extra conserved quantities were derived so that this work could be used to further investigate the integrability of \Eref{TodaS-iso}. 

\section{The stochastic Toda lattice with dissipation}\label{dissipative-toda}

Recall that after introducing noise in the Toda lattice in section \Sref{stochastic-toda}, we observed that the total energy of the lattice was not bounded in the infinite time limit. 
We thus want to introduce a nonlinear dissipation term in the Toda lattice to bound the energy, while remaining compatible with the geometrical structure used to add noise, that is to preserve the tri-diagonal structure of the $L$ matrix.  
Such nonlinear dissipation can be constructed using a double bracket structure from the coadjoint motion of the equation. 

\subsection{The double bracket dissipation}
\label{Dissipation}

The double bracket dissipation was first introduced by \cite{brockett1991dynamical} in the context of diagonalizations of matrices in linear programming and shortly after by \cite{bloch1992completely} for a description of the Toda lattice different from the one used in this work.
Indeed, they used the fact that the flow of the open Toda lattice has the property to diagonalise the $L$ matrix for large time. 
This state corresponds to all the particles at infinity with constant speed (the eigenvalues of $L$) and no interactions. 

The concept of a double bracket dissipation was then further extended to mechanical systems in \cite{bloch1996euler} and more recently by \cite{gaybalmaz2013selective,gaybalmaz2014geometric} where both energy and Casimir dissipations were studied and applied to fluid mechanical systems of Lie-Poisson form, that is which dynamics restricted to a coadjoint orbit, although infinite dimensional.  
We will first review this dissipative term and show how a slight modification of it can be implemented in the Toda lattice in the next section. 

For a dynamical system on a coadjoint orbit with Casimir $C$ (invariant function on the coadjoint orbit), the general dissipative term of double bracket form was introduced by \cite{gaybalmaz2013selective, gaybalmaz2014geometric} to give the dissipative Euler-Poincar\'e equation 
\begin{equation}        
    \frac{d}{dt} \mu + \mathrm{ad}^*_\frac{\partial h}{\partial \mu} \mu 
    + \theta\, \mathrm{ad}^*_\frac{\partial C}{\partial \mu} \left [ \frac{\partial C}{\partial \mu}, \frac{\partial h}{\partial \mu} \right ]^\flat = 0 \,,
    \label{EP-Diss}
\end{equation}
where the isomorphism $\flat : \mathfrak g\to \mathfrak g^*$ is associated to a given pairing, that can in general be different from the natural pairing of the Lie algebra $\mathfrak g$ where $\mu\in \mathfrak g^*$. 
The parameter $\theta$ controls the amount of dissipation and is an inverse time scale, the time scale for the system to relax to one of its minimum energy positions. 
Notice that for infinite dimensional systems, variational derivatives must be used instead of partial derivatives. 
The energy, or Hamiltonian $h(\mu)$ of this system decays as 
\begin{equation}
    \frac{d}{dt}h(\mu) = - \,\theta \left \| \left [ \frac{\partial C}{\partial \mu},\frac{\partial h}{\partial \mu} \right ]  \right \|^2\, ,
    \label{SD-diss-SD}
\end{equation}
where the norm is associated to the pairing used in the definition of the flat operation. 
This equation shows that the system is forced towards a position where the right-hand side vanishes, which is a condition compatible with an equilibrium solution of the original deterministic system. 
We refer to \cite{gaybalmaz2014geometric} for an extensive discussion on this condition. 

\subsection{The dissipative Toda lattice}

For the Toda lattice in Flaschka's variables, we will need a different term as Casimir functions are not directly available for this system.
We will instead use the term parametrized by $\theta\in \mathbb R$
\begin{equation}
    \dot L = \mathrm{ad}^{R,*}_LL + \theta \mathrm{ad}^{R,*}_{\mathrm{ad}^{R,*}_{L}L}L
    \label{Toda-diss}
\end{equation}
and show that it preserves the coadjoint orbit and is a dissipative term.
Notice that we do not need to convert the second $\mathrm{ad}^*$ operator in the double bracket from the dual of the Lie algebra to the Lie algebra as it is a symmetric matrix and the isomorphism $\flat:\mathfrak g^*\to \mathfrak g$ is the transpose operation. 

First, the Lie-Poisson formulation of \Eref{Toda-diss} is
\begin{equation}
  \dot F(L) = \{F,H_1\}_R - \theta\, \left  \langle \mathrm{ad}^{R,*}_{\frac{\partial H_1}{\partial L}} \frac{\partial H_1}{\partial L}, \mathrm{ad}^*_{\frac{\partial F}{\partial L}}\frac{\partial H_1}{\partial L}\right \rangle\, ,
    \label{Toda-diss-LP}
\end{equation}
where the pairing is the trace pairing, and the first bracket is the $R$-Lie-Poisson bracket of \Eref{R-LP-bracket}.
Indeed, we just compute 
\begin{equation*}
    \langle \nabla F ,\mathrm{ad}^*_{\mathrm{ad}^*_{L}L}L \rangle = \left \langle \mathrm{ad}_{\mathrm{ad}^*_{L}L}\frac{\partial F}{\partial L} ,L \right \rangle = -\left \langle \mathrm{ad}^*_{L}L ,\mathrm{ad}^*_{\frac{\partial F}{\partial L}}L \right \rangle\, , 
\end{equation*}
and use $\frac{\partial H_1}{\partial L} =L$ to get \Eref{Toda-diss-LP}. 
This form of the double bracket is different from the one described in the previous section as the adjoint action is used instead of the coadjoint action.

In term of the variables $a_i,b_i$, the dissipative Toda lattice becomes
\begin{eqnarray}
        \dot b_i &= b_i(a_i -a_{i+1})+ 2\theta b_i (b_{i-1}^2 - 2b_i ^2 +b_{i+1}^2)\\
        \dot a_i &= 2(b_{i-1}^2-b_{i}^2) +  2\theta \left ( b_i^2( a_{i+1} - a_i)- b_{i-1}^2 ( a_i-a_{i-1}  ) \right )\, ,
    \label{Toda-diss-ab}
\end{eqnarray}
where as for the noise terms, no equation for the other diagonals of $L$ exist. 
Indeed, if the coadjoint orbit where not conserved by this flow, we would need to include extra variables for the outer-diagonals of $L$ which would have equations coupled with these two.
Written in this form, the new terms have a clear meaning of dissipation as they both behave as discrete nonlinear second order operators as we will see in more detail in \Sref{continuum-limit}.

From the equation of motion in the Lie-Poisson form in \Eref{Toda-diss-LP}, we can compute the time evolution of any integral of motion to get
This formula is a direct consequence of 
\begin{equation}
 \frac{d}{dt} \frac{1}{i+1}\mathrm{Tr}(L^{i+1}) = - \theta \langle \mathrm{ad}^*_{\nabla H} \nabla H, \mathrm{ad}^*_{\nabla H_i}\nabla H\rangle\, ,
    \label{Hi-diss}
\end{equation}
which does not necessarily mean that the right hand side is strictly negative (for $\theta>0$), unless we pick the Hamiltonian of the Toda flow. 
Indeed, the equation becomes 
\begin{equation}
 \frac{d}{dt}H_1 = - \theta \|\mathrm{ad}^*_{\nabla H} \nabla H_1\|^2\, ,
    \label{H2-diss}
\end{equation}
and right-hand side is strictly negative, which shows that the energy decays in time. 
In physical coordinates, this formula reads
    \begin{equation}
        \frac{d}{dt}H_1 = -2\theta\sum_i\left ( (y_{i}-y_{i+1})^2 e^{x_{i+1}-x_{i}}+ (e^{x_{i+1}-x_{i}}- e^{x_{i}-x_{i-1}})^2\right )\, .
        \label{E-decay}
    \end{equation}
This dissipative Toda lattice will then tend to a stationary position $(x_i^s(t),y_i^s)$ where the right hand side of \Eref{E-decay} vanishes, that is 
\begin{equation*}
    \sum_i(y^s_{i}-y^s_{i+1})^2 e^{x^s_{i+1}-x^s_{i}}+ \sum_i (e^{x_{i+1}^s-x_{i}^s}- e^{x_{i}^s-x_{i-1}^s})^2= 0\, . 
\end{equation*}
The solution of this positive definite quantity is if $x^s_i(t)= y t$ and thus $y^s_i=y$ for some values of $x$ and $y$, or $x^s_i(t)-x^s_{i+1}(t)= \Delta x$ with $x_i^s(t)= x_i(0) + y t$ for some $\Delta x$ and then $y^s_i=y$.
In the first case the condition $y^s_i=y$ comes from the fact that the particles must remain at the same position, thus have the same speed. 
In the second case, the first term of the previous equation vanishes only if $y^s_i=y^s_{i+1}$ for all $i$.
The value of $y$ can be found from the conservation of total momentum, even in this dissipative case, but finding the value $x$ from the initial conditions for periodic conditions is difficult and not interesting for this work. 
In the open Toda lattice, the first and last particles are at $-\infty$ and $\infty$ and the equilibrium solution is $x=\pm \infty$. 

Finally, by combining the noise and the dissipation, the Toda lattice in physical coordinates has the following equation of motion
\begin{eqnarray}
    \dot x_i &= y_i + \frac12 \theta (e^{x_i-x_{i-1}}-2e^{x_{i+1}-x_i}+ e^{x_{i+2}-x_{i+1}})\\
    dy_i &=   \left ( e^{x_i-x_{i-1}} - e^{x_{i+1}-x_i}\right )dt + \nonumber \\
    &+2\sigma e^{(x_i-x_{i-1})/2}\circ dW_t^{i-1} - 2\sigma e^{(x_{i+1}-x_i)/2}\circ dW_t^i\nonumber \\
    &+\frac12 \theta \left ( ( y_{i+1}-y_i) e^{x_{i+1}-x_i} - ( y_i-y_{i-1})  e^{x_i-x_{i-1}} \right )dt \,  .
    \label{TodaS-diss}
\end{eqnarray}
The Fokker-Planck equation of this equation can be computed, but the stationary distribution seems not to be a Gibbs distribution (that is the exponential of the negative of the energy), as the noise is only a degenerate stochastic Hamiltonian system, that is it does not appear in the position equation, which would have been needed to obtain a Gibbs distribution, see \cite{arnaudon2016noise}. 

\subsection{Continuum limit}\label{continuum-limit}

One of the interesting features of the Toda lattice is that in the limit of an infinite number of particles, or the continuum limit, the lattice converges to Burger equation or the KdV equation if higher order terms are included in the approximation. 
The precise limit can be subtle for the Toda lattice as shocks can form and the Burger's model will no longer be valid after a shock. 
This was extensively studied by \cite{deift1998continuum}, where various after-shocks scenarios were considered. 
Here, we will derive the continuum limit of the dissipative stochastic Toda lattice, and, as the dissipation will prevent the formation of shocks we will not consider such scenarios. 

We will follow \cite{deift1998continuum} for the derivation of the continuum limit by defining the continuous functions interpolating through the discrete system to be 
\begin{equation}
    a(\epsilon k,t) = a_k(t)\qquad \mathrm{and}\qquad b(\epsilon k,t) = b_k(t)\, ,
\end{equation}
where $\epsilon$ corresponds to the distance between two particles at rest.
We then rescale the time as $t\to t/\epsilon$ and the diffusion coefficient as $\theta \to \theta/\epsilon$ to obtain from the dissipative Toda \Eref{Toda-diss-ab}, the continuous equations
\begin{eqnarray}
     \dot a &=  2\left (b^2(1+ \theta\, a_x)\right )_x\\
    \dot b &= b (a_x+ 2\theta\, b_{xx})\, , 
\end{eqnarray}
In the non-dissipative case, setting $a=2b$ recovers the inviscid Burger's equation, but in this dissipative case, no such reduction seem to exist. 
It is interesting to notice that a double bracket dissipation constructed with the $R$-matrix has a simple and well-defined continuous limit to a rather unusual nonlinear dissipation. 
Indeed, the momentum $\int a dx$ is conserved, but the energy is affected by the dissipation as 
\begin{equation}
  \frac{d}{dt} E = \frac{d}{dt} \frac12 \int \left ( \frac12 a^2 + b^2\right ) dx = \theta \int  b^2\left (2 b_{xx} - a_x^2\right ) dx\, . 
\end{equation}
The simplest condition for the right hand side to vanish is if $2b_{xx} = a_x^2$, which does not seem to have any physical meaning.
Further investigation of the properties of this nonlinear dissipation could be interesting and especially the form of the travelling wave solutions, which are analytically difficult to compute. 

To compute the continuum limit of the stochastic Toda lattice, we first assume an isotropic noise and rescale it as $\sigma_i= e_i \sigma/\sqrt{\epsilon}$ for all $i$, where $e_i$ has only $1$ in the $i$'s component of the first diagonal. 
We then obtain the following continuum approximation of the complete stochastic dissipative Toda lattice
\begin{eqnarray}
    \dot b &= b (a_x+ 2\theta b_{xx})\\
    d a &=  2\left (b^2(1+ \theta a_x)\right )_x\, dt + 2 \sigma b_x  dW_{t,x}\, ,
    \label{ab-sto}
\end{eqnarray}
where the noise is white in space and time, or delta correlated in space and time, that is $\mathbb E[W_{x,t} ,W_{x',t'}] = \delta(x-x') \delta (t-t')$. 
Notice that again, the stochastic term is in It\^o form, but is equivalent to a Stratonovich integration. 
We will not rigorously prove that this stochastic PDE is the continuum limit of the stochastic Toda lattice as this will be beyond the scope of this investigation, but we only want to verify that this stochastic PDE has a similar noise as the discrete system by comparing the auto-covariance function of the stochastic processes.
We can formally show that the auto-covariance of the discrete system \Eref{TodaS-diss} converges to the auto-covariance of the continuous model \Eref{ab-sto} by first computing  
\begin{eqnarray*}
    \mathrm{Cov}(a_k,a_{k'})&= \frac{ 4\sigma^2}{ \epsilon} \mathbb E \left ( \int (b_{k-1} dW_t^{k-1} - b_k dW_t^k) \int (b_{k'-1} dW_t^{k'-1} - b_{k'} dW_t^{k'}) \right )\\
    &=  \frac{4\sigma^2}{ \epsilon^2} \mathbb E \int ( b_k( \delta_{k,k'} b_{k'}-  \delta_{k,k'-1}  b_{k'-1})\\
    &\hspace{20mm}-b_{k-1}(\delta_{k-1,k'} b_{k'} -\delta_{k-1,k'-1} b_{k'-1}  )) \,  dt\, ,
\end{eqnarray*}
where used the fact that the Wiener processes are not correlated. 
We also used the It\^o isometry, given for a process $\mu_t$
\begin{equation}
    \mathbb E\left (\left (\int \mu_t dW\right)^2\right) = \mathbb E\left ( \int \mu_t^2 dt\right)\, . 
    \label{ito-isometry}
\end{equation}
In the limit $\epsilon\to0$, we then arrive at 
\begin{eqnarray*}
    \lim_{\epsilon\to 0} \mathrm{Cov}(a_k,a_{k'})&= 4 \sigma^2 \mathbb E \left ( \int \delta(x-x')b_x(x) b_{x'}(x') dt \right )\\
    &= \mathrm{Cov}(a(x),a(x'))\, ,
\end{eqnarray*}
where the last step is computed from the process \Eref{ab-sto}.
This calculation suggests that the stochastic dissipative Toda lattice converges to the continuous process \Eref{ab-sto} almost surely in the limit $\epsilon\to 0$. 
In general, the other choices of $\sigma_i$ could give any spatial correlation of the noise. 
On can for example only have a single Wiener process, and $\sigma_i$ non-zero for several particles, such that in the limit we can have a noise of the form $\sigma(x) dW_t$, where $\sigma(x)$ is the limiting function.

\section{Conclusion}

In this paper, we implemented a stochastic deformation and the double bracket dissipation in the integrable Toda lattice while preserving the coadjoint orbit structure of the original system. 
This makes these deformations non-standard, and possibly new, with still a connection to physical variables, even if the original motivation has little to do with physical phenomena. 
Indeed, stochastic perturbation are often additive, and dissipation linear to model interaction with a heat bath for example, but here, both are nonlinear. 
This analogy with the Langevin equation suggests that more can be done in the direction of statistical physics, and in particular deriving the stationary distribution, or understanding the interaction of noise and dissipation together with the original nonlinearities of the Toda lattice. 
Such investigations would rely on detailed numerical simulations that we did not undertake in this work. 
Other investigations with numerical simulations as well as analytical consideration may include the study of the $3$-particle Toda lattice, where one would expect, due to its low dimensionality, to directly investigate the existence of random attractor, in the context of random dynamical systems, \cite{arnaudon2016noise,arnold1995random}. 

Another intriguing finding is the existence of a well-defined continuum limit of the dissipation and of the noise terms, even if they were derived from a setting far from any notion of continuous limit. 
They both have a rather unusual form and may deserve more investigations which are out of the scope of this work. 
In particular for the dissipative Toda lattice on the existence and form of travelling waves (which in the limit $\theta\to 0$ tends to a shock), as compared with the viscid Burger equation for example. 
Also, it would be interesting to see if any generalisation of the Cole-Hopf transformation exists for this equation, which would allow it to be mapped to a possible two-dimensional heat equation. 
For the stochastic PDE with dissipation, many problems are open in the realm of stochastic analysis which we leave for future works, but it is interesting to remark that any spatial correlation of the noise can be used and remain compatible with the finite dimensional system. 

Finally, another interesting line of future research is the question of complete integrability of the stochastic Toda lattice, without dissipation. 
The stochastic term is different from being of the form of a higher order Toda flow multiplying a Wiener process (unless a single process is used with constant amplitude for all particles).  
Instead, the stochastic term has a different Wiener process for each particle, which makes the system non-integrable.  
In fact, we cannot recast the geometric stochastic deformation of this system into a stochastic isospectral problem unless one of the stochastic terms is discarded. 
In this case, the coadjoint orbit is not preserved, and a tri-diagonal initial condition will quickly fill the entire symmetric matrix, but we can write a stochastic iso-spectral problem for this system.  
A question then remains, which is if we can explicitly solve this stochastic spectral problem on the space of full symmetric matrices.  

\section*{Acknowledgements}
The author wants to thank A. Castro, D. Holm, T. Ratiu, C. Tomei, W. Pan, J. Gibbons and A. Hone for helpful discussions during the course of this work. 
The author is supported by the EPSRC through award EP/N014529/1 funding the EPSRC Centre for Mathematics of Precision Healthcare.
\section*{References}

\bibliographystyle{iopart-num.bst} 
\bibliography{biblio.bib,biblio-thesis.bib}

\end{document}